\documentclass[%
reprint,
superscriptaddress,
tbtags,
amsmath,amssymb,
aps,
pre,
floatfix,
]{revtex4-2}

\usepackage[colorlinks=true, citecolor=blue]{hyperref}
\usepackage{graphicx}
\usepackage{lmodern}
\usepackage{xfrac}
\usepackage{booktabs}
\usepackage{tabularray}
\UseTblrLibrary{booktabs}
\usepackage{longtable}
\usepackage{scalerel,stackengine}
\usepackage{xcolor}

\newcommand\pig[1]{\scalerel*[5pt]{\big#1}{%
  \ensurestackMath{\addstackgap[1.5pt]{\big#1}}}}
\newcommand\pigl[1]{\mathopen{\pig{#1}}}
\newcommand\pigr[1]{\mathclose{\pig{#1}}}
\newcommand{\Var}{\mathrm{Var}}

\DeclareMathOperator{\csch}{csch}

\begin{document}
\title{Tightening the Thermodynamic Uncertainty Relations with Null-entropy Events: What we learn when nothing happens}
\author{Abhaya S. Hegde}
\email{a.hegde@rochester.edu}
\affiliation{
Department of Physics and Astronomy, University of Rochester, Rochester, New York 14627, USA
}
\affiliation{University of Rochester Center for Coherence and Quantum Science, Rochester, New York 14627, USA}
\author{Andr\'e M. Timpanaro}
\affiliation{Universidade Federal do ABC, 09210-580 Santo Andr\'e, Brazil}
\author{Gabriel T. Landi}
\affiliation{
Department of Physics and Astronomy, University of Rochester, Rochester, New York 14627, USA
}
\affiliation{University of Rochester Center for Coherence and Quantum Science, Rochester, New York 14627, USA}

\begin{abstract}
Fluctuation theorems establish that thermodynamic processes at the microscale can occasionally result in negative entropy production. 
At such scales, another distinct possibility becomes more likely: processes in which no entropy is produced overall.
In this work, we explore the constraints imposed by such null-entropy events on the fluctuations of thermodynamic currents.
By incorporating the probability of null-entropy events, we obtain tighter bounds on finite-time thermodynamic uncertainty relations derived from fluctuation theorems.
Our results are directly applicable to both quantum and classical systems that satisfy the fluctuation theorem.
We validate our framework using an example of a qudit SWAP engine.
\end{abstract}

\maketitle

\section{Introduction}
Non-equilibrium processes inherently produce irreversible entropy~\cite{Callen85, Munster70}. 
At the microscopic scale, thermal fluctuations can induce large variability in individual realizations that may transiently violate average thermodynamic trends, including momentary reversals of heat and particle flow. 
Such events are vanishingly unlikely in macroscopic systems, revealing a behavior that sharply contrasts with the smooth, deterministic evolution predicted by classical thermodynamics~\cite{Evans94,Gallavotti95, Gallavotti952}.
Stochastic thermodynamics offers a rigorous framework for describing this regime by treating entropy production as a fluctuating observable defined along the individual trajectories of a stochastic process~\cite{Seifert2008,Esposito09,Campisi11,seifertstochastic2012,Esposito12}. This trajectory-level formulation shifts the focus from macroscopic averages to single realizations, where the statistics of entropy production becomes the key quantity.

At the heart of stochastic thermodynamics lie the celebrated fluctuation theorems~\cite{Jarzynksi97,Crooks1998,Crooks99,Evans02,Wang02,Jarzynski04,Seifert05,Harris07,Campisi11,jarzynski_2013,Campisi14}, which enforce fundamental constraints on entropy fluctuations. 
These universally exact relations reveal the extent to which microscopic systems can transiently defy macroscopic thermodynamic expectations. If 
$P(\Sigma,\phi)$ denotes the joint distribution of entropy production $\Sigma$ and some thermodynamic current $\phi$ (e.g. heat or work) in the forward process, and $P_B(-\Sigma, -\phi)$, the corresponding distribution for backward dynamics, then the fluctuation theorems imply (with $k_B = 1$) 
\begin{equation}
\label{eq:FT}
\frac{P_B(-\Sigma, -\phi)}{P(\Sigma, \phi)} = e^{-\Sigma}.
\end{equation}
Entropy-reducing fluctuations are thus exponentially suppressed. 
Applying Jensen's inequality, we recover the second law of thermodynamics expressed in terms of the average entropy production: $\langle \Sigma \rangle \geq 0$.

While fluctuation theorems capture the full spectrum of entropy production, they provide limited insight into events that leave no thermodynamic trace.
This is the focus of the present paper: stochastic events for which no entropy is produced, i.e., $\Sigma = 0$, henceforth referred to as null-entropy events.
Such events are a hallmark of microscopic machines operating under stochastic dynamics. 
These may correspond to true inactivity, or to sequences of events whose effects cancel out, leaving no net energy or particle exchange. 
For these events, the standard fluctuation theorem, which primarily addresses nonzero dissipation, offers no new intuition. 
Nevertheless, the prevalence of null-entropy events suggests that the structure of $P(\Sigma=0)$, the probability of zero entropy production, can still carry meaningful thermodynamic information. 
In this work, we attend to this often-overlooked question: what can be learned when seemingly nothing happens?

To that end, we explore another family of fundamental results related to entropy production called the thermodynamic uncertainty relations (TURs)~\cite{SeifertTUR15, England16, horowitz2017proof, pietzonka2017finite, Seifert18, seifertstochastic2018, barato_unifying_2019}.
These relations declare constraints on the signal-to-noise ratio of thermodynamic currents in terms of the average entropy production.
TURs can be broadly categorized into two classes. 
Steady-state TURs~\cite{SeifertTUR15,Seifert16,England16,Esposito16,horowitz2017proof,Seifert18,Landi19,Saulo25} apply to long-time integrated currents, common in autonomous machines or steady-state heat transport settings. 
In contrast, finite-time TURs~\cite{pietzonka2017finite,hasegawa2019fluctuation,Timpa19,vanvu2022quantum,Guarnieri23,Hasegawa24} address the fundamental precision-dissipation trade-offs that govern transient dynamics, such as those encountered in individual strokes of a heat engine.
For long-time integrated currents, the probability that $\Sigma = 0$ diminishes to be vanishingly small.
Therefore, this paper focuses exclusively on the finite-time formulations, as this is the regime where the notion of null-entropy events remains meaningful.

\medskip

Several forms of TURs have been formulated to address systems operating over finite timescales. To distinguish them, it is useful to first specify the backward process characterized by $P_B$ in Eq.~\eqref{eq:FT}.
Within the framework of the exchange fluctuation theorem~\cite{Jarzynski04}, or strong Galavotti-Cohen symmetry~\cite{Gallavotti952}, the backward process coincides with the forward process ($P_B = P$). 
For such cases, the TURs have the general form 
\begin{equation}
\label{eq:tur}
\frac{{\rm Var}(\phi)}{\langle \phi \rangle^2} \geqslant f\pigl(\langle \Sigma \rangle \pigr),
\end{equation}
where $f(x)$ is a monotonically decreasing function, such as $2/(e^{x} - 1)$~\cite{hasegawa2019fluctuation, proesmans2017}, or an even tighter version $\csch^2[g(x/2)]$ with $g^{-1}(x) = x\tanh(x)$~\cite{Timpa19}.
Here, $\phi$ is some thermodynamic current, e.g., the work extracted from a single cycle of a heat engine. 
The bound therefore establishes a constraint on the possible fluctuations of $\phi$ relative to its mean.
Consequently, TURs reveal that enhanced precision in $\phi$, or equivalently reduced fluctuations, inevitably leads to a higher entropy production.
Extensions to cases where $P_B \neq P$ are deferred to Sec.~\ref{sec:asym_TUR}. 

Considering Eq.~\eqref{eq:tur} for now, a natural question is whether the knowledge of $p_0 \equiv P(\Sigma=0)$ results in an improved bound for TURs. 
In this work, we answer that in the affirmative.
Our main result is, with both $p_0$ and $\langle \Sigma\rangle$ available, the bounds of Eq.~\eqref{eq:tur} are replaced by 
\begin{equation}
\label{eq:new_tur}
\frac{{\rm Var}(\phi)}{\langle \phi \rangle^2} \geqslant \frac{1}{1-p_0} f\Biggl(\frac{\langle \Sigma \rangle}{1-p_0}\Biggr) + \frac{p_0}{1-p_0} \equiv f\bigl(\langle \Sigma \rangle, p_0\bigr).
\end{equation}
These new bounds are tighter than those in Eq.~\eqref{eq:tur} for non-zero $p_0$.
The knowledge of null‑entropy events imposes stricter limits on the possible fluctuations of thermodynamic currents.
Remarkably, this improvement in the TUR bound applies to both classical and quantum systems, provided they satisfy the fluctuation theorem given in Eq.~\eqref{eq:FT}.

This paper is organized as follows. 
We begin with a minimal toy model in Sec.~\ref{sec:minimal_example} to illustrate the properties of $p_0$ 
and establish its relation to fluctuations.
By utilizing the knowledge of $p_0$, we discuss a modified fluctuation theorem in Sec.~\ref{sec:MFT} and derive new bounds on TURs in Sec.~\ref{sec:MTUR}.
We identify null‑entropy events in a bipartite quantum system subjected to a two-point measurement scheme in Sec.\ref{sec:tpm}.
The improvements to TUR bounds are demonstrated with a qudit SWAP engine example in Sec.~\ref{sec:examples}. 
We briefly mention extensions to asymmetric versions of TURs in Sec.~\ref{sec:asym_TUR} and conclude with a discussion in Sec.~\ref{sec:Disc}.

\section{\label{sec:minimal_example}Minimal example model for bounds on null-entropy events}

To motivate the importance of null-entropy events, we consider a minimal toy model in which the system can produce some positive entropy ($\Sigma=+\sigma$) with probability $p_+$, consume said entropy ($\Sigma=-\sigma$) with probability $p_-$, or do nothing ($\Sigma = 0$) with probability $p_0$. 
The fluctuation theorem from Eq.~\eqref{eq:FT} implies $p_- = p_+ e^{-\sigma}$. 
Together with normalization, we have
\begin{equation}
\label{eq:norm}
    p_0 + p_+^{}\bigl(1 + e^{-\sigma}\bigr) = 1.
\end{equation}
The average entropy production is 
$\langle \Sigma \rangle = p_+^{} \sigma \bigl(1 - e^{-\sigma}\bigr)$.
Combined with Eq.~\eqref{eq:norm}, this implies that 
\begin{equation}
\label{eq:ave_ep}
\langle \Sigma \rangle = \sigma(1-p_0)\tanh\biggl(\frac{\sigma}{2}\biggr).
\end{equation}
In principle, $p_0,p_+,p_-$ and $\sigma$ are all the independent variables necessary to fully determine the statistics of entropy production $\Sigma$. 
However, on account of normalization and the fluctuation theorem, only two independent parameters are sufficient: $\sigma$, which represents the magnitude of entropy change during active events, and $p_0$, the probability associated with null-entropy events.
These not only uniquely determine the average entropy production, as shown in Eq.~\eqref{eq:ave_ep}, but also govern the fluctuations since $\langle \Sigma^2 \rangle = \sigma^2 (p_++p_-) = \sigma^2 (1-p_0)$. 
Consequently, the noise-to-signal ratio (NSR) reads
\begin{equation}\label{eq:nsr_ep}
    \frac{{\rm Var}(\Sigma)}{\langle \Sigma \rangle^2} 
    = \frac{1}{(1-p_0)} \frac{1}{\tanh^2(\sigma/2)}-1.
\end{equation}

\begin{figure}
    \centering
    \includegraphics[width=0.85\linewidth]{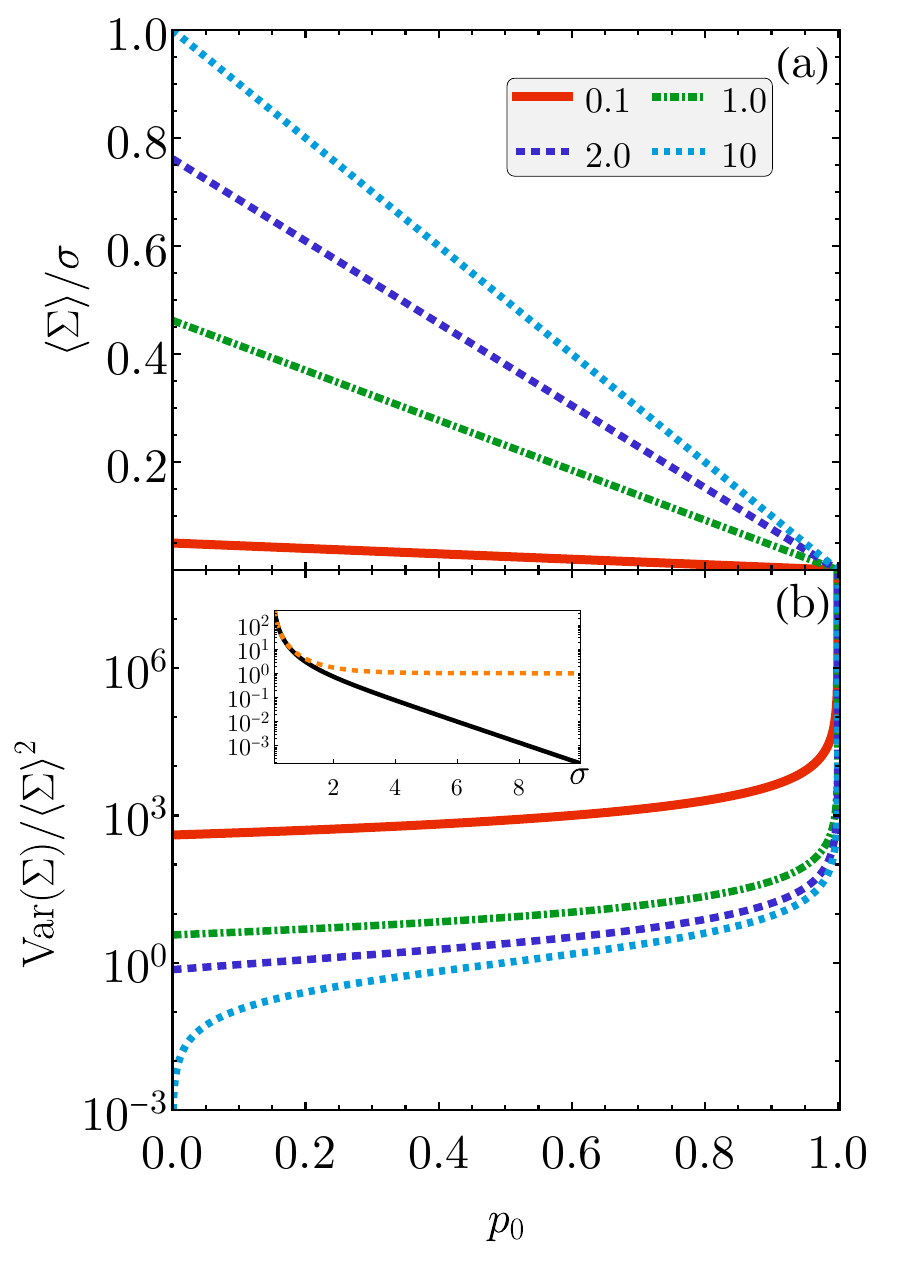}
    \caption{{\bf (a)} Average entropy production $\langle \Sigma\rangle$ and {\bf (b)} noise-to-signal ratio ${\rm Var}(\Sigma)/\langle \Sigma\rangle^2$ for the minimal model discussed in Sec.~\ref{sec:minimal_example} [Eqs.~\eqref{eq:ave_ep} and \eqref{eq:nsr_ep}].
    The quantities are plotted as a function of the null-entropy probability $p_0$ for different values of $\sigma$, as shown in (a).
    Significant fluctuations arise precisely in regimes dominated by null-entropy events.
    {\bf Inset of (b)}: the functions $[\tanh^2(\sigma/2)-1]^{-1}$ (solid black) and $[\tanh^2(\sigma/2)]^{-1}$ (orange dashed).
    }
    \label{fig:min_model}
\end{figure}

Equations~\eqref{eq:ave_ep} and~\eqref{eq:nsr_ep} are plotted in Fig.~\ref{fig:min_model} as a function of $p_0$ for different values of $\sigma$. 
When null-entropy events are rare ($p_0\ll 1$), the NSR is approximately 
\begin{equation}
    \frac{{\rm Var}(\Sigma)}{\langle \Sigma \rangle^2} \simeq  \frac{1}{\tanh^2(\sigma/2)}-1.
\end{equation}
This is plotted in black in the inset of Fig.~\ref{fig:min_model}(b) demonstrating that, in this regime, fluctuations are substantial only for processes with minimal entropy production. 
In the opposite limit, as $p_0 \simeq 1$, where null-entropy events are frequent, the NSR~\eqref{eq:nsr_ep} diverges for any $\sigma$.
This divergence scales as $1/(1-p_0)$ and indicates a universal amplification of relative fluctuations when reversible-like trajectories dominate, independent of the typical entropy production.
The corresponding prefactor is now only $1/\tanh^2(\sigma/2)$ which is now lower-bounded (shown in orange in the inset), so that fluctuations never vanish.
Because this is a consequence of the fluctuation theorem itself, this observation remains insensitive to microscopic details.
In essence, fluctuations are directly linked to the frequency of null-entropy events, irrespective of the model.

\section{\label{sec:MFT}Modified fluctuation theorem}

We now reconsider the general fluctuation theorem~\eqref{eq:FT}, described in terms of the joint probability $P(\Sigma,\phi)$ of the entropy production and some generic thermodynamic current $\phi$. 
Without loss of generality, we  assume null-entropy events to satisfy both $\Sigma = 0$ and $\phi = 0$, and denote the corresponding probability by $p_0$; any thermodynamic current can be redefined to vanish when $\Sigma = 0$. 
From Eq.~\eqref{eq:FT}, null-entropy events occur with equal probability in forward and backward processes, $P(0,0) = P_B(0,0) = p_0$.

Given a certain process characterized by a probability distribution $P(\Sigma,\phi)$, we now examine an alternative, fictitious process which excludes the null-entropy events but is otherwise identical to the original. 
Specifically, for $p_0 \ne 0$, we consider a process characterized by a distribution $\widetilde{P}(\Sigma, \phi)$ that satisfies $\widetilde{P}(0,0) = 0$ and 
\begin{equation}
    \widetilde{P}(\Sigma, \phi) = P(\Sigma, \phi)/(1-p_0),
\end{equation}
where the factor $1-p_0$ ensures that $\widetilde{P}$ remains normalized. 
We identically define a corresponding backward process characterized by $\widetilde{P}_B$.
Rewriting the fluctuation theorem~\eqref{eq:FT} for this modified process then yields
\begin{equation}
\label{eq:nonzero_FT}
\frac{\widetilde{P}_{B}(\Sigma, \phi)}{\widetilde{P}(-\Sigma, -\phi)} = e^{\Sigma}.
\end{equation}
This shows that the hypothetical process described by the auxiliary distribution $\widetilde{P}$ also obeys the same fluctuation theorem. 

A property we shall repeatedly employ is that for any function $g(\Sigma, \phi)$, expectation values can be decomposed as follows:
\begin{equation}
\label{eq:expectations_main_rule}
    \pigl\langle g\bigl(\Sigma, \phi\bigr) \pigr\rangle = p_0 g(0, 0) + (1 - p_0) \pigl\langle g\bigl(\Sigma, \phi\bigr) \pigr\rangle^{\!\sim},
\end{equation}
where $\langle\cdots\rangle^{\!\sim}$ is an expectation over the modified process characterized by the distribution $\widetilde{P}$.
Particularly, for calculating the moments of $\Sigma$ and $\phi$, the first term vanishes as $g(0,0)\equiv 0$, and we are simply left with
\begin{equation}
    \pigl\langle g\bigl(\Sigma, \phi\bigr) \pigr\rangle =  (1 - p_0) \pigl\langle g\bigl(\Sigma, \phi\bigr) \pigr\rangle^{\!\sim}.
\end{equation}
On the other hand, for the integral fluctuation theorem $\langle e^{-\Sigma}\rangle = 1$, which is the Jarzynski equality derived from the detailed fluctuation theorem in Eq.~\eqref{eq:FT}, we find
\begin{equation}
    1=\langle e^{-\Sigma}\rangle = p_0 + (1-p_0) \langle e^{-\Sigma}\rangle^{\!\sim}.
\end{equation}
It then follows that the process characterized by $\widetilde{P}$ also satisfies the Jarzynski equality: 
\begin{equation}
    \bigl\langle e^{-\Sigma} \bigr\rangle = 1\quad \to \quad \bigl\langle e^{-\Sigma}\bigr\rangle^{\!\sim} = 1,
\end{equation}
which, of course, is consistent with Eq.~\eqref{eq:nonzero_FT}.

\section{\label{sec:MTUR}Modified bounds for TUR}

We proceed now to the TURs in Eq.~\eqref{eq:tur} where $f(\langle \Sigma\rangle)$ is a monotonically decreasing function in $\langle \Sigma \rangle$.
Isolating the null-entropy contributions, we rewrite the noise-to-signal ratio of any thermodynamic current using Eq.~\eqref{eq:expectations_main_rule} as
\begin{equation}
\label{eq:new_precision}
\frac{{\rm Var}(\phi)}{\langle \phi \rangle^2} = \frac{p_0}{1 - p_0}+ \frac{1}{1-p_0}\frac{{\rm Var}(\phi)^{\!\sim}}{\langle \phi \rangle^{\!\sim 2}} .
\end{equation}
By the non-negativity of the variance, ${\rm Var}(\phi)^{\!\sim}\geqslant 0$, it immediately follows that 
\begin{equation}\label{eq:hasegawa}
\frac{{\rm Var}(\phi)}{\langle \phi \rangle^2} \geqslant \frac{p_0}{1-p_0} = \left(\frac{1}{p_0} - 1\right)^{-1}.
\end{equation}
This is precisely the bound introduced by Hasegawa in Ref.~\cite{Hasegawa24}.
Notably, it illustrates how $p_0$ places fundamental constraints on the fluctuations of finite-time processes. 

The bound can be further sharpened by noting that the fictitious process $\widetilde{P}$ also obeys its own TURs due to fluctuation relations in Eq.~\eqref{eq:nonzero_FT}, 
\begin{equation}
    \frac{{\rm Var}(\phi)^{\!\sim}}{\langle \phi \rangle^{\!\sim 2}} \geqslant f\big(\langle \Sigma\rangle^{\!\sim}\big) = f\biggl(
    \frac{\langle \Sigma\rangle}{1-p_0}
    \biggr).
\end{equation}
Plugging this in Eq.~\eqref{eq:new_precision} then leads to our main result presented in Eq.~\eqref{eq:new_tur}. The resulting bound retains the form of the original TUR, but with a different right-hand-side function:
\begin{equation}\label{eq:new_f}
    f\pigl(\langle \Sigma \rangle, p_0\pigr) = \frac{1}{1-p_0} f\Biggl(\frac{\langle \Sigma \rangle}{1-p_0}\Biggr) + \frac{p_0}{1-p_0} .
\end{equation}
It represents a modified bound for finite processes that incorporates both the average entropy production and the null-entropy probability. It is thus tighter than both the original TURs~\eqref{eq:tur} and the Hasegawa bound~\eqref{eq:hasegawa}.

\begin{figure}
    \centering \includegraphics[width=0.9\linewidth]{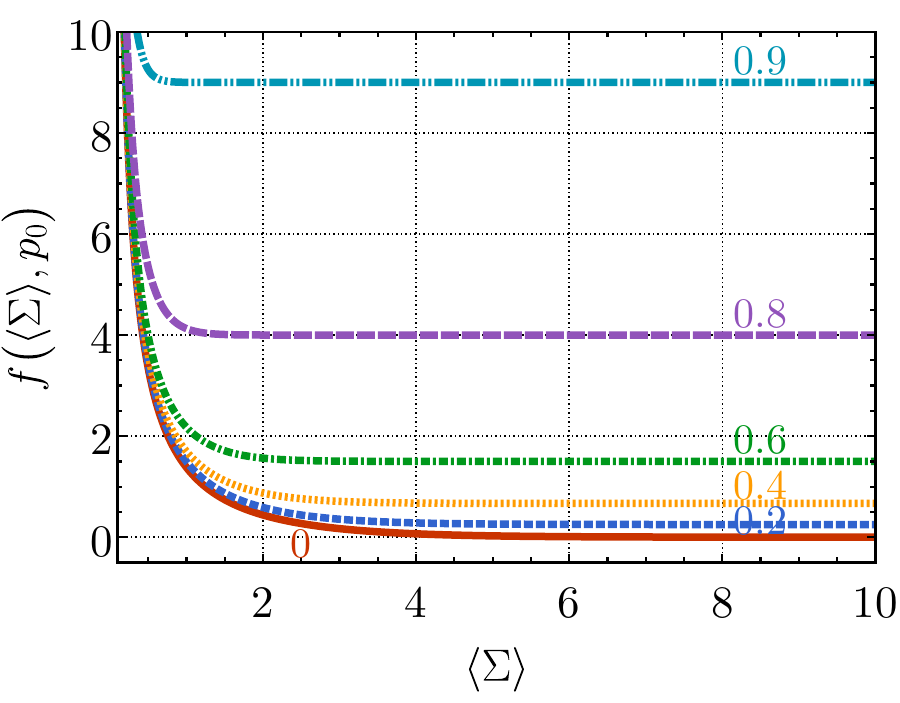}
    \caption{
    Modified \emph{TUR-de-force} bounds from Eqs.~\eqref{eq:new_f} and \eqref{eq:tur-de-force} shown for different values of the probability $p_0$ associated with zero integrated current and entropy production. The bound gets tighter as $p_0$ increases.
    }
    \label{fig:new_TUR}
\end{figure}

To illustrate this result, we focus on the particular choice 
\begin{equation}
\label{eq:tur-de-force}
f\bigl(\langle \Sigma \rangle\bigr) = \csch^2\bigl[g\bigl(\langle \Sigma \rangle/2\bigr)\bigr], \quad g^{-1}(x) = x \tanh(x).
\end{equation}
This bound, termed \emph{TUR-de-force} in Ref.~\cite{Timpa19}, was shown to be the tightest (and saturable) among all bounds that (i) satisfy an exchange fluctuation theorem and (ii) depend only on $\langle \Sigma \rangle$ on the right-hand side.
In Fig.~\ref{fig:new_TUR}, we plot Eq.~\eqref{eq:new_f} as a function of $\langle \Sigma\rangle$ for different values of $p_0$. 
In the limit $p_0\to 0$ we recover the usual TUR since Eq.~\eqref{eq:new_f} satisfies
\begin{equation}
\label{eq:old_f}
    f\pigl(\langle\Sigma\rangle,0\pigr) = f\bigl(\langle\Sigma\rangle\bigr).
\end{equation}
For $p_0\neq 0$, however, we obtain 
\begin{equation}
    f\pigl(\langle \Sigma\rangle,p_0\pigr) \geqslant f\bigl(\langle \Sigma\rangle\bigr), \qquad \forall p_0.
\end{equation}
The difference thus quantifies the extent to which knowledge of $p_0$ tightens the fluctuations. 
This result is not specific to \emph{TUR-de-force}, but also holds for 
\begin{equation}
f\bigl(\langle \Sigma \rangle\bigr) = \frac{2}{e^{\langle \Sigma \rangle} - 1},
\end{equation}
another bound derived from the fluctuation relations~\cite{hasegawa2019fluctuation}. Remarkably, this was the same bound found in discrete time-symmetric setting~\cite{proesmans2017}. 
By direct inspection for the bounds considered here, $\partial f\bigl(\langle \Sigma\rangle,p_0\bigr)/\partial p_0 \geqslant 0$ for all $p_0 \in [0,1)$.
In fact, for any non-negative function $f\bigl(\langle \Sigma\rangle\bigr)$ that monotonically decreases with $\langle \Sigma\rangle$ and satisfies 
$\partial\bigl[\langle\Sigma\rangle\, f\bigl(\langle\Sigma\rangle\bigr)\bigr]/\partial \langle \Sigma \rangle > -1$, the corresponding function $f\bigl(\langle \Sigma\rangle,p_0\bigr)$ defined in Eq.~\eqref{eq:new_f} \emph{increases} monotonically in $p_0$ at fixed $\langle \Sigma\rangle$.
Consequently, for such functions, the knowledge of $p_0$ therefore can only tighten the bound. 
On the other hand, if the precision is known, this fact can be used to establish an upper bound for $p_0$.

The improved bound can also be understood in light of the probabilistic structure presented in Sec.~\ref{sec:minimal_example}.
There, the entropy production statistics are represented by a three-point distribution $\{-\sigma, 0, \sigma\}$, in which the null-entropy probability $p_0$ being non-zero introduces an additional support point absent in a purely two-point setting.
The non-zero contributions $\Sigma\neq 0$ retain the same two-point structure at $\pm\sigma$ for which the optimal distribution minimizing the variance at fixed $\langle \Sigma\rangle$ was first identified in Ref.~\cite{Timpa19}.
This distinction clarifies why the presence of null-entropy events enables a strictly tighter TUR than is possible in the conventional two-point case.

It is worth considering the implications to the thermodynamic uncertainty theorem (TUT)~\cite{Guarnieri23}, which follows from the fluctuation theorem~\eqref{eq:FT} when $P_B = P$. 
The TUT captures the contribution to precision from higher moments of $\Sigma$ by replacing the right hand side in Eq.~\eqref{eq:tur} from $f\bigl(\langle\Sigma\rangle\bigr)$ with the average of a function of entropy production, i.e. $\langle f(\Sigma) \rangle$. Specifically, it states
\begin{equation}
\label{eq:TUT}
\frac{\Var(\phi)}{\langle \phi \rangle^2} \geqslant \frac{1}{\langle\tanh(\Sigma/2)\rangle} - 1.
\end{equation}
Crucially, this formulation does not lend itself to any changes with the incorporation of newly acquired knowledge of $p_0$.
To see this, we apply TUT to the auxiliary distribution $\widetilde{P}$, yielding
${\rm Var}(\phi)^{\sim}/\langle \phi \rangle^{\sim 2} \geqslant 1/\langle \tanh(\Sigma/ 2) \rangle^{\sim} - 1$.
Now, substituting this in Eq.~\eqref{eq:new_precision} and recognizing $(1 - p_0) \langle \tanh(\Sigma / 2) \rangle^{\sim} = \langle \tanh(\Sigma / 2) \rangle$, the bound reduces to the original expression~\eqref{eq:TUT}. 

\section{\label{sec:tpm}Null-entropy events in a two-point measurement scheme}

To provide some intuition for what null-entropy events physically mean, we consider a scenario involving work and heat exchange in a bipartite system undergoing unitary dynamics. 
We use the two-point measurement (TPM) protocol~\cite{Lutz07,Esposito09,Campisi11} to analyze the  trajectory‑level statistics of two quantum systems $A$ and $B$, prepared in their own thermal states and driven by a time-dependent drive. 
The Hamiltonian during the protocol is $H(t) = H_A(t) + H_B(t) + V(t)$, where $V(t)$ describes their interaction.
For simplicity, we assume that $V(t)$ vanishes at the beginning and end of the protocol, so the initial and final Hamiltonians read 
\begin{equation}
H_i = H_A(0) + H_B(0), \qquad H_f = H_A(\tau) + H_B(\tau). 
\end{equation}

At $t=0$, the joint system is initialized in the product state
$\rho_0 = \rho_A^{\rm th} \otimes \rho_B^{\rm th}$ where
$\rho^{\rm th}_{\alpha}=e^{-\beta_\alpha H_{\alpha}(0)}/Z_{\alpha}(0)$ 
are thermal states at inverse temperatures
$\beta_\alpha$, and $Z_{\alpha}(0)={\rm tr}\bigl\{e^{-\beta_\alpha H_{\alpha}(0)}\bigr\}$ for $\alpha = \{A, B\}$.
Let $H_\alpha(t)|i_{\alpha}(t)\rangle = E_{i_{\alpha}}(t) |i_{\alpha}(t)\rangle$ denote the instantaneous eigenstates of the local Hamiltonians. 
The TPM scheme starts with a local projective energy measurement on the joint system basis $|i (0)\rangle \equiv |i_{A} (0),i_{B} (0)\rangle$.
The system then evolves unitarily under the Hamiltonian $H(t)$ for duration $\tau$ which can be represented by a unitary 
$U(\tau, 0) = \mathcal{T} \exp\bigl[{\int_0^\tau dt \, H(t)}\bigr]$, where $\mathcal{T}$ is the time-ordering operator. 
At time $\tau$, we again measure the system which yields $|f(\tau) \rangle \equiv |f_{A} (\tau),f_{B} (\tau)\rangle$.
The change in energy of each subsystem, at the stochastic level, is then 
$\Delta E_\alpha = E_{f_{\alpha}}(\tau)-E_{i_{\alpha}}(0)$.

We shall now define the thermodynamic variables of interest---heat ($Q_\alpha$), work ($W$), and entropy production ($\Sigma$)---for each stochastic trajectory as follows:
\begin{equation}
\label{eq:tpm_stats}
Q_\alpha = -\Delta E_\alpha,
\hspace{0.3em}
W=\sum_{\alpha}\Delta E_\alpha, 
\hspace{0.3em}
\Sigma=\sum_\alpha\beta_\alpha (\Delta E_\alpha-\Delta F_\alpha),
\end{equation}
where $\Delta F_\alpha = - \beta_\alpha^{-1} \log[Z_{\alpha} (\tau)/Z_{\alpha} (0)]$ is the difference in equilibrium free energy of each system. 
According to the TPM protocol, the distribution for entropy production is
\begin{equation}\label{eq:P_sigma_TPM}
\begin{aligned}
    P(\Sigma) = & \sum_{i,f} p_{i}^{\rm th}\,
    \pigl|\pigl\langle f(\tau) \pigr|U(\tau, 0)\pigl|i(0)\pigr\rangle\pigr|^2 \\[-0.1cm]
     &\qquad  \times \delta\Biggl[\Sigma - \sum_\alpha\beta_\alpha (\Delta E_\alpha-\Delta F_\alpha)\Biggr],
\end{aligned}    
\end{equation}
where $p_i^{\rm th} = p_{i_A}^{\rm th}p_{i_B}^{\rm th}$ where each term is given by $p^{\rm th}_{i_{\alpha}}= e^{-\beta_\alpha E_{i_{\alpha}}(0)}/Z_{\alpha}(0)$  
as the probability of measuring the eigenstate
$|i_{\alpha}(0)\rangle$.
For convenience, we display $\delta(a-b)$ to represent the Kronecker delta $\delta_{a,b}$.
Distributions of other thermodynamic quantities can be obtained similarly by replacing the argument of the $\delta$-function in Eq.~\eqref{eq:P_sigma_TPM} with the appropriate combination of energy changes, such as in Eq.~\eqref{eq:tpm_stats}.
Hence, this procedure naturally extends to joint distributions, such as 
$P(Q_\alpha, W)$, and more general multi-variable cases.

We now define the following dephasing superoperator, acting on an arbitrary operator $\mathcal{O}$ as, 
\begin{equation}\label{eq:dephasing_map}
\begin{aligned}
\mathcal{D}_{\Sigma}(\mathcal{O}) =& 
\sum_{i, f}\, \pigl|i(0)\pigr\rangle\pigl\langle i(0)\pigl|\, \mathcal{O} \,\pigr| f(\tau)\pigr\rangle  \pigl\langle f(\tau)\pigr| \\
&\ \times 
\delta\Biggl[\sum_\alpha\beta_\alpha (\Delta E_\alpha-\Delta F_\alpha)\Biggr].
\end{aligned}
\end{equation}
This operator $\mathcal{D}$ projects out the off-diagonal elements of $\mathcal{O}$ between subspaces with nonzero entropy production.
The probability that a trajectory produces no entropy, i.e., $p_0 \equiv P(\Sigma = 0)$, is then given by:
\begin{equation}\label{eq:p0_tpm}
    p_0 = {\rm tr}\pigl\{D_{\Sigma}\bigl(U^\dagger\bigr) U \rho^{\rm th}_A \rho^{\rm th}_B\pigr\}.
\end{equation}
This result connects null-entropy events to the overlap between a forward unitary $U$ and the dephased version of its backward unitary $U^\dagger$.
For a reversible process, $\mathcal{D}_\Sigma(U^\dagger) = U^\dagger$ and $p_0 = 1$. 

Equation \eqref{eq:p0_tpm} shows that null-entropy events are linked specifically to the energetic coherences generated by the unitary.
However, only those coherences connecting the subspaces with different entropy production are relevant.
The interpretation simplifies when the initial and final Hamiltonians coincide: the systems are driven out of equilibrium, but their Hamiltonians return to their original form at the end of the protocol, so $\Delta F_\alpha=0$ and $\bigl|f(\tau)\bigr\rangle = \bigl|f(0)\bigr\rangle$ rendering initial and final bases the same.
The only constraint left in the dephasing map~\eqref{eq:dephasing_map} is $\delta\bigl(\beta_A \Delta E_A + \beta_B \Delta E_B\bigr)$.

If $\beta_A = \beta_B$, the null-entropy events satisfy $\Delta E_A + \Delta E_B = 0$. The dephasing occurs only within the energetic coherence of $H_{A}(0) + H_{B}(0)$, so that only degenerate subspaces are preserved.
For $\beta_A \ne \beta_B$, the constraint typically enforces $\Delta E_A = 0 = \Delta E_B$.
While there can be suitably engineered cases with $\Delta E_{\alpha} \ne 0$ but still $\sum_{\alpha} \beta_\alpha \Delta E_\alpha = 0$, these vanish under arbitrarily small changes of $\beta_\alpha$.
Thus, null-entropy events correspond to no energy change in either system individually.
The dephasing map then factorizes into independent dephasings in the eigenbases of $H_{A}(0)$ and $H_{B}(0)$, removing coherences even within degenerate subspaces.

\begin{figure*}
    \centering
    \includegraphics[width=\linewidth]{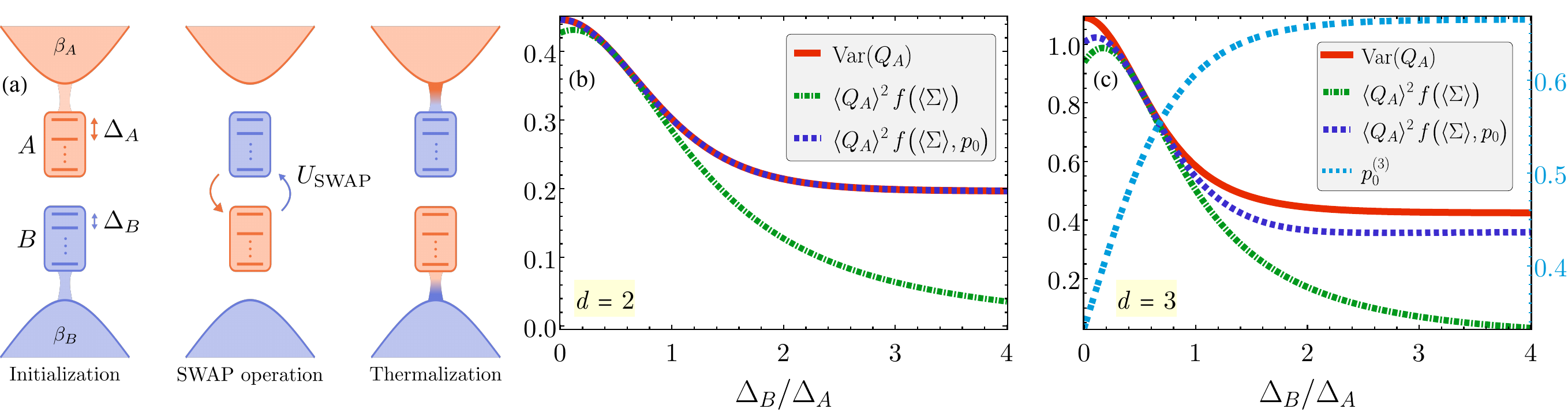}
    \caption{{\bf (a) -- (c)}: Analysis of fluctuations in heat extracted for a qudit SWAP engine. 
    {\bf (a)} Schematic of a $d$-level qudit SWAP engine. Two equally spaced $d$-level qudits act as the working medium and are connected to a hot and cold reservoirs. Qudits initiate in a thermalized state, after which a SWAP unitary extracts work from them, ending the cycle with re-thermalization.
    Fluctuations in the SWAP engine (solid red), and \emph{TUR-de-force} bounds with (dashed blue) and without (dot-dashed green) the knowledge of null-entropy values $p_0$ [Eqs.~\eqref{eq:new_f}--\eqref{eq:old_f}], are presented for {\bf (b)} qubits with $d = 2$, and {\bf (c)} qutrits with $d = 3$. The null-entropy values are particularly useful for qubits, as they directly determine the variance, whereas for qutrits, they serve as a means to enhance existing bounds.
    These plots assume values $\beta_A = 1$, $\beta_B = 2$.}
    \label{fig:SWAP_engine}
\end{figure*}

\section{\label{sec:examples}Example: SWAP engine}
Motivated by the result from Sec.~\ref{sec:tpm}, we utilize the TPM framework in a two qudit SWAP engine~\cite{Sacchi21} to illustrate the improvements to precision. A paradigmatic model of a quantum thermal machine is provided by a simpler two qubit version in Refs.~\cite{Timpa19,Quan2007}. 
The engine operates through repeated unitary interactions involving two $d$-level qudits, denoted $A$ and $B$, which serve as the working medium [see Fig.~\ref{fig:SWAP_engine}\,(a)]. Qudits $A$ and $B$, with level spacings $\Delta_{A}$ and $\Delta_{B}$, couple respectively to a hot reservoir at inverse temperature $\beta_{A}$ and a cold reservoir at $\beta_{B}$. Both qudits are assumed to have uniform level spacings so that their free Hamiltonian appears as $H_\alpha = \Delta_\alpha \sum_{m=0}^{d-1} m |m\rangle \langle m|$ for $\alpha = \{A, B\}$. We fix $\hbar = 1$, $\Delta_A > \Delta_B$ and $\beta_A < \beta_B$ without loss of generality.

Each engine cycle proceeds in two steps: (i) a \emph{reset} step, where each qudit of the working medium is brought into contact with their corresponding reservoir and instantaneously thermalized to the corresponding Gibbs state $\rho^{\mathrm{th}}_{\alpha} = e^{-\beta_\alpha H_\alpha}/Z_\alpha$ where $Z_\alpha = {\rm tr}\bigl\{e^{-\beta_\alpha H_\alpha}\bigr\}$; and (ii) a \emph{unitary interaction} step, where the two qudits undergo a SWAP operation described by the unitary in the computational basis $\{|m\rangle\}_{0}^{d-1}$,
\begin{equation}
\label{eq:swap_unitary}
U_{\rm SWAP} = \sum_{n, m = 0}^{d-1}|n\rangle \langle m| \otimes |m \rangle \langle n|.
\end{equation}
This operation exchanges the energy of the qudits, resetting them to their new equilibrium states determined by their respective baths.
A sequential implementation of these two strokes repeatedly facilitates the operation of this thermal machine.
For each cycle, the heat released by the hot bath relates to the energy change in qudit $A$, $Q_A = -\Delta E_A$, and similarly $Q_B = -\Delta E_B$ is released into the cold reservoir. The work from the unitary interaction is then $W = -(Q_A + Q_B) = \Delta E_A + \Delta E_B$. The entropy production per cycle follows $\Sigma = -\beta_A Q_A - \beta_B Q_B$.
The joint probability $P(\Sigma, Q_A, W)$ was computed analytically in Ref.~\cite{Sacchi21} for arbitrary dimension $d$. 
Naturally, it satisfies the exchange fluctuation theorem 
\begin{equation}
\label{eq:swap_eft}
    \frac{P(-\Sigma, -Q_A, -W)}{P(\Sigma, Q_A, W)} = e^{-\Sigma}.
\end{equation}
Here, we focus on the mean and variance of $Q_A$. 
Let $C_\alpha^{(d)} = d \coth\left(d \beta_\alpha \Delta_\alpha/2\right)$ and adopt the notation $C_\alpha \equiv C_\alpha^{(1)}$. 
Then from Ref.~\cite{Sacchi21},
\begin{subequations}
\begin{align}
\langle Q_A \rangle =
& \ \frac{\Delta_A}{2} \left[ C_A^{\phantom{(}} - C_A^{(d)} - \left(C_B^{\phantom{(}} - C_B^{(d)}\right) \right], \\
\Var(Q_A) =
&\ \frac{\Delta_A^2}{2}\biggl[d^2 - 1 + C_A^2 + C_B^2 - C_A^{\phantom{(}} C_B^{\phantom{(}} \\
& - d\pigl(C_A^{} - C_B^{}\pigr)\pigl(C_A^{(d)} - C_B^{(d)}\pigr) - d^2 C_A^{(d)} C_B^{(d)}\biggr].\notag
\end{align}
\end{subequations}
The average work extracted and entropy produced in each cycle are simply
\begin{align}
\label{eq:swap_work}
\langle W \rangle &= \frac{\Delta_B - \Delta_A}{\Delta_A} \langle Q_A \rangle, \\
\label{eq:swap_ent}
\langle \Sigma \rangle &= (\beta_B - \beta_A) \langle Q_A \rangle + \beta_B \langle W \rangle.
\end{align}

The null-entropy probability for this example follows from Eq.~\eqref{eq:p0_tpm}.
For an initial state $|i\rangle = |n, m\rangle$ and final state $|f\rangle = |n', m'\rangle$, the dephasing map~\eqref{eq:dephasing_map} that disconnects null-entropy subspace from others takes the form,
\begin{equation}
\begin{aligned}
    \mathcal{D}_{\Sigma}(\mathcal{O}) =&\ \sum_{\substack{n, m \\ n', m'}}^{d - 1}\bigl|n, m\bigr\rangle \bigl\langle n, m\bigr| \mathcal{O} \bigl | n', m' \bigr \rangle \bigl \langle n', m' \bigr|\\
    &\quad \times \delta\Bigl(E_{n'_{\!A}} - E_{n^{}_{\!A}}\Bigr) \ \delta\Bigl(E_{m'_{\!A}} - E_{m^{}_{\!A}}\Bigr).
\end{aligned}
\end{equation}
For $U^{\dagger}_{\rm SWAP}$ available from Eq.~\eqref{eq:swap_unitary}, the action of the dephasing map appears simply as,
\begin{equation}
    D_{\Sigma}\bigl(U^\dagger_{\rm SWAP}\bigr) = \sum_{n = 0}^{d-1} \bigl| n, n\bigr\rangle \bigl\langle n, n\bigr|,
\end{equation}
which decoheres all off-diagonal elements.
Since both qudits begin in their thermal states, we have 
\begin{equation}
\label{eq:swap_p0}
p_0^{(d)} =  \sum_{n=0}^{d-1} \frac{e^{-n\beta_A \Delta_A}}{Z_A} \frac{e^{-n\beta_B \Delta_B}}{Z_B}.
\end{equation}

The simplest case of $d = 2$ recovers a qubit SWAP engine. Upon fixing the average entropy produced, this example is reminiscent of the minimal model discussed in Sec.~\ref{sec:minimal_example}.
Estimating the variance for such an example requires three independent quantities represented in $\Var(Q_A) = \Sigma^2 (1 - p_0) - \langle \Sigma \rangle^2$.
As noted in Eq.~\eqref{eq:swap_eft}, since this satisfies exchange fluctuation theorem, we can apply the \emph{TUR-de-force} from Eq.~\eqref{eq:tur-de-force} as the tightest version of TUR.
Reference~\cite{Timpa19} presented an approach to optimize variance, and thereby TUR bounds, by considering only two support points: the entropy production per event $\Sigma$ and the average entropy produced $\langle \Sigma \rangle$. Introducing an additional independent support, $p_0$, should align precisely with the variance, thereby saturating the bound. This behavior is clearly demonstrated in our example, as shown in Fig.~\ref{fig:SWAP_engine}\,(b), where the variance and previously established bounds are compared.

Extending this to higher dimensions is straightforward.
For $d = 3$, a tighter TUR is established from Eq.~\eqref{eq:new_tur}. In Fig.~\ref{fig:SWAP_engine}\,(c), we plot the variance, \emph{TUR-de-force}, and our modified version with $p_0$ against the energy spacing ratio of the qutrits.
Across the parameter ranges, the new bound proves to be consistently tighter than the \emph{TUR-de-force}.
The tightness of the new bound depends on the relative magnitude of $p_0$, which is plotted on the second y-axis in Fig.~\ref{fig:SWAP_engine}\,(c). A larger $p_0$ improves the bound, as evidenced by the decreasing gap between the variance and the new bound as $p_0$ increases.
Such observations can also be made for any higher $d$ (see Appendix~\ref{appsec:higher_d}).

\section{\label{sec:asym_TUR}Extensions to Asymmetric TURs}
We now address processes that lack a time-reversal symmetry, resulting in $P_B \ne P$.
This distinction between forward and backward processes is particularly relevant in the presence of measurement and feedback. 
Indeed, such processes exhibit a different constraint on the noise-to-signal ratio from the one prescribed in Eq.~\eqref{eq:tur}. This missing connection was first established by Potts and Samuelsson in Ref.~\cite{Potts19} as
\begin{equation}
\label{eq:potts_sam}
    \frac{{\rm Var}^{}(\phi) + {\rm Var}^{}(\phi)_B}{\pigl(\langle \phi \rangle^{} + \langle \phi \rangle_B^{} \pigr)^2} \geqslant \biggl[{\exp\biggl(\frac{\langle \Sigma \rangle + \langle \Sigma \rangle_B}{2}\biggr) - 1}\biggr]^{\!-1},
\end{equation}
where $\langle \cdots \rangle_B$ and ${\rm Var}(\cdots)_B$ denote mean and variance of thermodynamic current in the backward process.
Consequently, the signal-to-noise ratio in a given measurement can become arbitrarily large as long as the corresponding backward process compensates for it. 
Such efforts to extend TUR to asymmetric processes were further advanced in Ref.~\cite{Timpa24} to establish a family of TUR-like relations which follow from the fluctuation theorem of the form given in Eq.~\eqref{eq:FT}. 
This result allowed signal-to-noise ratios for a convex combination of forward and backward processes such that
\begin{equation}
\label{eq:timpa_turs}
\frac{\alpha {\rm Var}^{}(\phi) + (1 - \alpha) {\rm Var}(\phi)_B^{}}{\pigl(\langle \phi \rangle^{} + \langle \phi \rangle_B^{} \pigr)^2} \geqslant \alpha^2 \biggl[1 - \frac{1}{\omega(\Sigma)}\biggr]^{-1},
\end{equation}
for $0 \leq \alpha \leq 1$ and a function of entropy production, $\omega(\Sigma) = \pigl\langle \bigl(1 - \alpha + \alpha e^{-\Sigma}\bigr)^{-1} \pigr\rangle$.

In addition to the forward and backward processes being asymmetric, if one also has access to $p_0$, Potts-Samuelsson TUR is modified (see Appendix~\ref{appsec:asymmetric_TUR}) to
\begin{align}
\label{eq:modified_potts_sam}
\frac{{\rm Var}(\phi) + {\rm Var}(\phi)_B}{\pigl[\langle \phi \rangle + \langle \phi \rangle_B\pigr]^2} \geqslant & \ \frac{1}{1-p_0} \Biggl[\exp\biggl(\frac{\langle \Sigma \rangle + \langle \Sigma \rangle_B}{2(1-p_0)}\biggr)-1\Biggr]^{-1} \notag \\ &+
\frac{p_0}{1-p_0}\frac{\langle \phi \rangle^2 + \langle \phi \rangle_B^2}{\pigl[\langle \phi \rangle + \langle \phi \rangle_B^{}\pigr]^2}.
\end{align}
The above expression is consistent with the existing bound from Eq.~\eqref{eq:potts_sam} when $p_0 = 0$.
On the right-hand side, while the first term captures the role of dissipation, the second term can be seen as a correction purely due to the asymmetry in forward and backward probability distributions.
Since $e^x - 1 \approx x$ for small $x$, we have the following for small total dissipation:
\begin{equation*}
\frac{{\rm Var}(\phi) + {\rm Var}(\phi)_B}{\pigl[\langle \phi \rangle + \langle \phi \rangle_B\pigr]^2} \gtrsim 
\frac{2}{\langle\Sigma\rangle+\langle\Sigma\rangle_B} +
\frac{p_0}{1-p_0}\frac{\langle \phi \rangle^2 + \langle \phi \rangle_B^2}{\pigl[\langle \phi \rangle + \langle \phi \rangle_B^{}\pigr]^2}.
\end{equation*}
Consider the dimensionless factor in the last term,
\begin{equation}
R \equiv \frac{\langle \phi\rangle^2+\langle \phi\rangle_B^2}{\bigl[\langle \phi\rangle+\langle \phi\rangle_B\bigr]^2},
\end{equation}
whose range depends on the sign of the product $\langle \phi \rangle \cdot \langle \phi \rangle_B$.
For $\langle \phi \rangle \cdot \langle \phi \rangle_B \ge 0$, the Cauchy–Schwarz inequality gives $1/2 \le R \le 1$, with $R=1/2$ when $\langle \phi \rangle = \langle \phi \rangle_B$ and $R=1$ when either term vanishes.
If $\langle \phi \rangle , \langle \phi \rangle_B < 0$, then $R \ge 1$.
In the first case, the last term in Eq.~\eqref{eq:modified_potts_sam} is bounded above by $p_0 / (1 - p_0)$; in the second, it is bounded below by the same expression. Remarkably, this value is the same as the last term in Eq.~\eqref{eq:new_f}.

\section{\label{sec:Disc}Discussions and Conclusions}
The improvements to thermodynamic uncertainty relation derived here introduces a new perspective on precision and dissipation trade‑offs by explicitly accounting for null-entropy events, i.e., transitions that preserve informational state of the system. Leveraging this information, our central result shows that when the probability of such events $p_{0}>0$, the conventional TUR bounds can be substantially tightened for any system that obeys the fluctuation theorem, revealing a special structure between thermodynamic cost and precision. 
Previous extensions of the TUR have primarily targeted multiple currents or quantum formulations. 
Our approach supplements them because we decompose the trajectories into two subensembles: ``active" and ``null."
Crucially, we obtain an analytically transparent bound that smoothly reduces to the conventional TURs as $p_{0}\to 0$.
Our analysis also encompasses processes that lack symmetry in forward and backward dynamics.

It would be misleading to dismiss the probability of null-entropy events $p_0$ as merely another point in the full spectrum of entropy distribution.
Physically, these events do not contribute to dissipation, making them fundamentally different from other values. 
Furthermore, they introduce a special symmetry in the statistics of entropy production since $p_0$ is the only point where the forward and backward distributions necessarily coincide, regardless of the process. These observations explain why $p_0$
is unique; the statistical properties of entropy production exhibit features that are consistent whether or not these null-entropy events are included.

Our derivation relies on a strict separation of trajectories into zero and nonzero entropy increments; for processes with continuously distributed entropy changes this dichotomy may appear less sharp.
Moreover, experimental limitations such as instrumental noise can obscure true null events, compromising the practical value of $p_{0}$.
Nevertheless, the framework remains highly applicable to systems where entropy changes are predominantly discrete.

This work contributes to the understanding that TURs not only represent fundamental limits, but they could also reveal structured trade-offs between precision and dissipation in stochastic processes. 
The incorporation of null‐entropy statistics opens new avenues for thermodynamic inference in intermittent systems. 
Extending these conceptual frameworks to address quantum jump processes represents a key direction to gain special insights into microscopic machines.
It would be especially interesting to explore the interplay between null‐entropy events and information flows in feedback‐controlled devices. 
Specifically, the question of whether feedback protocols can be designed to selectively amplify null-entropy trajectories for enhancing reversible work extraction while minimizing dissipation remains a topic for future research.

\begin{acknowledgments}
This research is primarily supported by the U.S. Department of Energy (DOE), Office of Science, Basic Energy Sciences (BES) under Award No. DE-SC0025516.
\end{acknowledgments}

\emph{Data Availability:}
The data that support the findings of this article are openly available~\cite{hegde_2026_18670310}.

\appendix
\onecolumngrid
\section{\label{appsec:higher_d}Qudit SWAP engine for higher dimensions}
\begin{figure}[!h]
    \centering
    \includegraphics[width=0.65\linewidth]{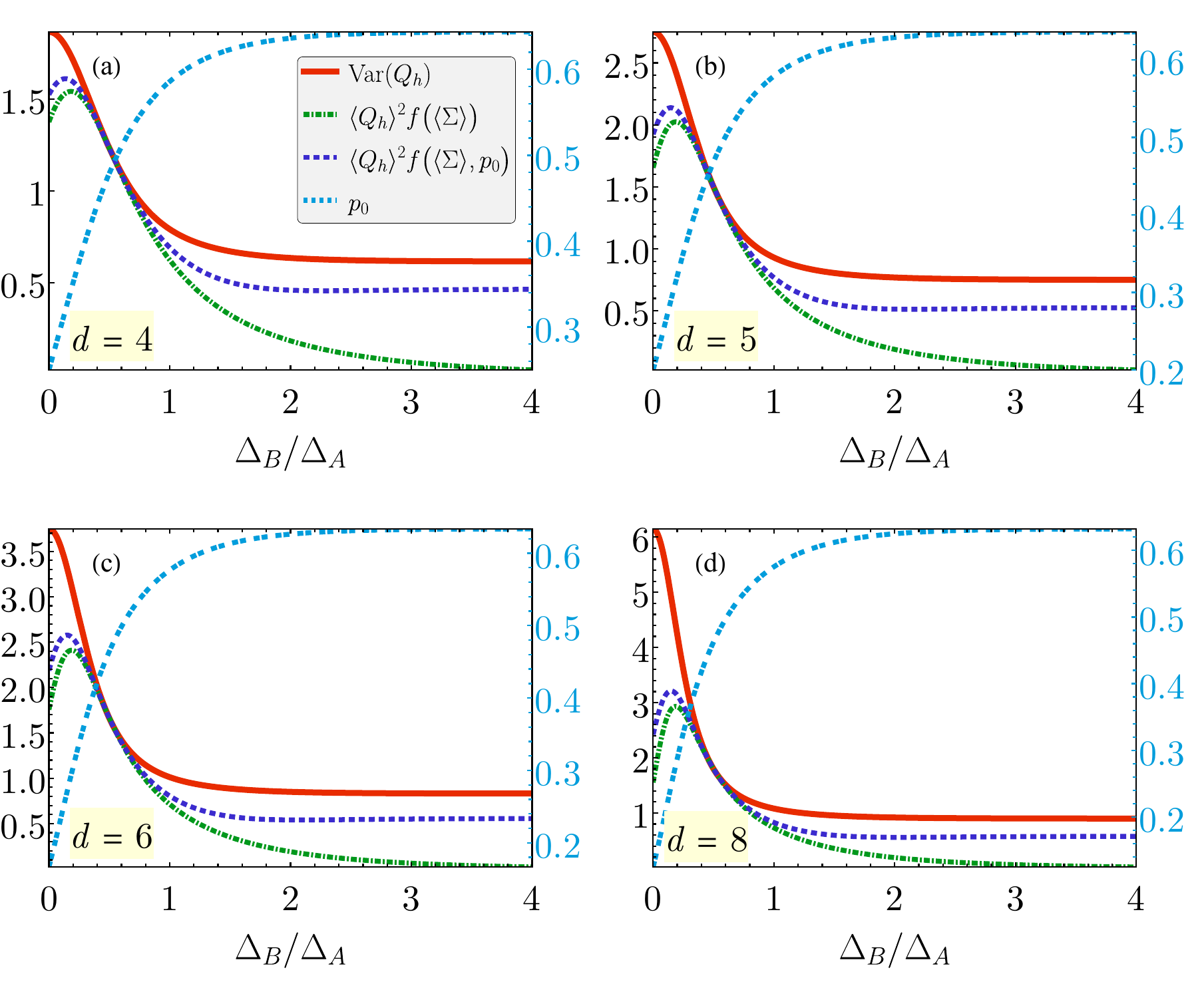}
    \caption{Extending the analysis done for the heat fluctuations in qudit SWAP engine [see Fig.~\ref{fig:SWAP_engine}] with {\bf (a)} $d = 4$, {\bf (b)} $d = 5$, {\bf (c)} $d = 6$, {\bf (d)} $d = 8$. Across all dimensions, the TUR bound tightens by supplying the null-entropy probability $p_0$. The extent to which $p_0$ varies is also plotted (in dashed light blue). These plots also assume the same parameters as in Fig.~\ref{fig:SWAP_engine}.}
    \label{appfig:higher_d}
\end{figure}

A similar analysis can be performed for qudits of higher dimensions (see Fig.~\ref{appfig:higher_d}).
In general, the expression for $p_0$ in a SWAP engine [Eq.~\eqref{eq:swap_p0}] can be simplified to
\begin{equation}
    p_0^{(d)} = 
    \frac{\pigl(1-e^{-\beta_A \Delta_A}\pigr)\pigl(1-e^{-\beta_B \Delta_B}\pigr)}
    {1-e^{-(\beta_A \Delta_A + \beta_B \Delta_B)}}
    \frac{1-e^{-d (\beta_A \Delta_A + \beta_B \Delta_B)}}
    {\pigl(1-e^{-d \beta_A \Delta_A}\pigr)\pigl(1-e^{-d \beta_B \Delta_B}\pigr)}.
\end{equation}
Thus, the range of $p_0^{(d)} \in [1/d, 1)$. 
A global minimum of $1/d$ is reached in the high-temperature limit as $\beta_A, \beta_B \ll 1$.
For low-temperatures $\beta_A, \beta_B \gg 1$, it approaches the upper bound of $1$ as expected.
For fixed level spacings and temperatures, $p_0^{(d)}$ strictly decreases with $d$. Subsequently, it saturates in the large-$d$ limit to,
\begin{equation}
p_0^{(d \to \infty)} = \frac{\pigl(1 - e^{-\beta_A \Delta_A}\pigr) \pigl(1 - e^{-\beta_B \Delta_B}\pigr)}
{1-e^{-\beta_A \Delta_A - \beta_B \Delta_B}}.
\end{equation}
This follows because, as the number of levels in each qudit grows, the null‑entropy transfer events become increasingly rare.
Consequently, the identification of null-entropy events enhances the precision to its utmost extent for qubits, while the magnitude of improvement diminishes as the dimensions increase.

\section{\label{appsec:asymmetric_TUR}Derivation of modified asymmetric TURs}
From the discussion in Sec.~\ref{sec:MFT}, it is clear that non-zero entropy and current variables with the distribution $\widetilde{P}$ obey their own fluctuation theorem [see Eq.~\eqref{eq:nonzero_FT}], which leads to a TUR of the following general form~\cite{Timpa24}:
\begin{equation}
\label{appeq:TUR_from_FUT}
\frac{\alpha {\rm Var}(\phi)^{\sim} + (1 - \alpha){\rm Var}(\phi)_B^{\sim}}{\pigl(\langle \phi \rangle^{\sim} + \langle \phi \rangle_B^{\sim}\pigr)^2} \geqslant F_\alpha\Bigl[\pigl\langle G_\alpha \bigl(\Sigma \bigr)\pigr\rangle^{\sim}\Bigr].
\end{equation}
Recall, that the $\langle \ldots \rangle^{\sim}$ and ${\rm Var}(\ldots)^{\sim}$ represent mean and variance over the auxiliary distribution $\widetilde{P}$. The notations without $\sim$ represent the actual distribution $P$.

We would like to rewrite the above equation for the original process characterized by $P$. To that end, we make use of the following substitutions obtained from Eq.~\eqref{eq:expectations_main_rule}:
\begin{subequations}
\label{eq:app_subs}
\begin{align}
\langle \phi \rangle &= (1- p_0) \langle \phi \rangle^{\sim}, \\
\Var(\phi) &= (1 - p_0) \Var(\phi)^{\sim} + p_0 (1 - p_0) \langle \phi \rangle^{\sim 2}.
\end{align} 
\end{subequations}
It then follows,
\begin{align}
\alpha \Var(\phi) + (1- \alpha) \Var(\phi)_B =& \ (1 - p_0) \bigl[\alpha \Var(\phi)^\sim + (1 - \alpha) \Var(\phi)_B^\sim\bigr] + p_0 (1 - p_0) \pigl(\alpha \langle \phi \rangle^{\sim 2} + (1 - \alpha) \langle \phi \rangle_B^{\sim 2}\pigr)\\[0.2cm]
\geqslant & \ (1 - p_0) \pigl( \langle \phi \rangle^{\sim} + \langle \phi \rangle_B^{\sim} \pigr)^{2}
F_\alpha\Bigl[\pigl\langle G_\alpha \bigl(\Sigma \bigr) \pigr\rangle^{\sim}\Bigr] + p_0 (1-p_0)\pigl(\alpha \langle \phi \rangle^{\sim 2} + (1 - \alpha)\langle \phi \rangle_B^{\sim 2}\pigr).
\end{align}
The inequality in the second line follows from Eq.~\eqref{appeq:TUR_from_FUT}.
Now, we shall eliminate the reference to auxiliary distribution using the simplifications from Eq.~\eqref{eq:app_subs}:
\begin{subequations}
\begin{align}
(1 - p_0) \pigl(\langle \phi \rangle^{\sim} + \langle \phi \rangle_B^{\sim} \pigr)^2 
&= \frac{1}{1-p_0} \pigl( \langle \phi \rangle + \langle \phi \rangle_B \pigr)^2 
\\[0.2cm]
p_0 (1-p_0) \Bigl(\alpha \langle \phi \rangle^{\sim 2} + (1 - \alpha)\langle \phi \rangle_B^{\sim 2}\Bigr) 
&= \dfrac{p_0}{1 - p_0}\Bigl(\alpha \langle \phi \rangle^2 + (1 - \alpha)\langle \phi \rangle_B^2\Bigr),
\\[0.2cm]
F_\alpha\Bigl[\pigl\langle G_\alpha \bigl(\Sigma \bigr) \pigr\rangle^{\sim}\Bigr] 
&= F_\alpha\biggl[ \frac{\langle G_\alpha(\Sigma) \rangle - p_0 G_\alpha(0)}{1-p_0} \biggr].
\end{align}
\end{subequations}
Substituting the above equations in Eq.~\eqref{appeq:TUR_from_FUT} results in the modified asymmetric TURs specifying the contributions from $p_0$ explicitly,
\begin{equation}
\label{appeq:modified_timpa_turs}
    \alpha {\rm Var}(\phi) + (1 - \alpha) {\rm Var}(\phi)_B^{} \geqslant \frac{p_0}{1-p_0}\Bigl( \alpha \langle \phi \rangle^2 + (1 - \alpha) \langle \phi \rangle_B^2 \Bigr) + \frac{\Bigl(\langle \phi \rangle + \langle \phi \rangle_B^{}\Bigr)^2}{1-p_0} F_\alpha\biggl[ \frac{\langle G_\alpha(\Sigma) \rangle - p_0 G_\alpha(0)}{1-p_0} \biggr],
\end{equation}
where $F_\alpha$ and $G_\alpha$ are functions of the average entropy production.

\subsection{Modifications to Potts-Samuelsson bound}
To obtain Potts-Samuelsson's symmetric TUR bound given in Eq.~\eqref{eq:potts_sam}, choose $\alpha = 1/2$, and notice that 
\begin{equation}
F_{\sfrac{1}{2}}\Bigl[\pigl\langle G_{\sfrac{1}{2}} \bigl(\Sigma \bigr) \pigr\rangle^{\sim}\Bigr] = \biggl[2\,\exp\biggl(\frac{\langle \phi \rangle^{\sim} + \langle \phi \rangle_B^{\sim}}{2}\biggr) - 1\biggr]^{-1}.
\end{equation}
This leads to the identification $F_{\sfrac{1}{2}}(x) = [2(e^x - 1)]^{-1}$.
To recognize $G_{\sfrac{1}{2}}(x)$, consider,
\begin{equation}
    \frac{\langle \Sigma \rangle^{\sim} + \langle \Sigma \rangle_B^{\sim}}{2} = \frac{\langle \Sigma \rangle^{\sim} + \pigl\langle - \Sigma e^{-\Sigma} \pigr\rangle^{\sim}}{2} = \biggl\langle \frac{\Sigma (1 - e^{-\Sigma})}{2} \biggr\rangle^{\sim}.
\end{equation}
This implies $G_{\sfrac{1}{2}}(x) = x(1-e^x)/2$. Hence, the bound from Eq.~\eqref{appeq:modified_timpa_turs} assumes the form mentioned in Eq.~\eqref{eq:modified_potts_sam}:
\begin{align}
{\rm Var}(\phi) +& {\rm Var}(\phi)_B \geqslant \frac{p_0}{1-p_0}\pigl( \langle \phi \rangle^2 + \langle \phi \rangle_B^2 \pigr) + \frac{\pigl(\langle \phi \rangle + \langle \phi \rangle_B\pigr)^2}{1-p_0} \biggl[\exp\biggl(\frac{\langle \Sigma \rangle + \langle \Sigma \rangle_B}{2(1-p_0)}\biggr)-1\biggr]^{-1}.
\end{align}

\subsection{Modifications to Timpanaro bounds}
A similar procedure to that mentioned above for a family of bounds from Ref.~\cite{Timpa24} [see Eq.~\eqref{eq:timpa_turs}] yields
\begin{equation}
F_\alpha(x) = \alpha^2 + \alpha \left( \frac{1}{x} - 1 \right ), \quad G_\alpha(x) = \frac{1 - e^{-x}}{1 - \alpha + \alpha e^{-x}},
\end{equation}
which can be substituted in Eq.~\eqref{appeq:modified_timpa_turs} to obtain new bounds with $p_0$.

\twocolumngrid

\bibliography{ref}

\end{document}